\documentclass[doublecol]{epl2}

\usepackage{tabularx}

\title{Fully gapped superconducting state in Au$_2$Pb: a natural candidate for topological superconductor}
\shorttitle{} 

\author{Y. J. Yu$^{1}$, Y. Xu$^{1}$, Y. Xing$^{2,3}$, J. Zhang$^1$, T. P. Ying$^1$, X. C. Hong$^1$, M. X. Wang$^1$, X. Zhang$^{2,3}$, S. Jia$^{2,3}$, J. Wang$^{2,3}$\footnote{E-mail: jianwangphysics@pku.edu.cn} \and S. Y. Li$^{1,4}$\footnote{E-mail: shiyan$\_$li@fudan.edu.cn}}
\shortauthor{Y. J. Yu\etal}

\institute
{$^1$ State Key Laboratory of Surface Physics, Department of Physics, and Laboratory of Advanced Materials, Fudan University, Shanghai 200433, China\\
 $^2$ International Center for Quantum Materials, School of Physics, Peking University, Beijing 100871, China\\
 $^3$ Collaborative Innovation Center of Quantum Matter, Beijing, China\\
 $^4$ Collaborative Innovation Center of Advanced Microstructures, Nanjing 210093, China}


\abstract
{We measured the ultra-low-temperature specific heat and thermal conductivity of Au$_2$Pb single crystal, a possible three-dimensional Dirac semimetal with a superconducting transition temperature $T_c \approx$ 1.05 K. The electronic specific heat can be fitted by a two-band $s$-wave model, which gives the gap amplitudes $\Delta_1$(0)/$k_BT_c$ = 1.38 and $\Delta_2$(0)/$k_BT_c$ = 5.25. From the thermal conductivity measurements, a negligible residual linear term $\kappa_0/T$ in zero field and a slow field dependence of $\kappa_0/T$ at low field are obtained. These results suggest that Au$_2$Pb has a fully gapped superconducting state in the bulk, which is a necessary condition for topological superconductor if Au$_2$Pb is indeed one.}

\begin{document}

\maketitle

\section{Introduction}

In recent years, the topological materials have attracted much attention because of their novel quantum states \cite{Hasan,Zhang,Bansil}. Among those topological materials, the topological superconductors (TSCs) are of particular interests. The TSCs have a full gap in the bulk and Majorana fermion states on the surface \cite{Zhang}. The TSCs are very important since they possess potential application to topological quantum computation due to the non-Abelian statistics of Majorana fermions \cite{Nayak}. Experimentally, there are several routes to obtain a TSC candidate. The first one is to artificially fabricate topological insulator/superconductor heterostructures \cite{Xu,Sun}. The second one is to dope a topological material. For example, the Cu-intercalated Bi$_2$Se$_3$ shows superconductivity below 3.8 K \cite{Hor}, which was considered as a TSC candidate \cite{Wray,Kriener,Sasaki,Matano}. The third one is to pressurize a topological material. Pressure-induced superconductivity was observed in some topological insulators, such as Bi$_2$Se$_3$, Bi$_2$Te$_3$, and Sb$_2$Te$_3$ \cite{Zhang2,Kirshenbaum,Zhu}. Superconductivity was also found in three-dimensional Dirac semimetal Cd$_3$As$_2$ under a tip or hydropressure \cite{Sheet,Wang3,He}. However, to confirm a TSC, it is essential to identify the Majorana fermion states on the surface. Unfortunately, no consensus has been reached on this important issue yet \cite{Fu}.

Besides these routes, in some rare cases, a natural TSC candidate may exist if a stoichiometric superconductor manifests topological properties under ambient pressure. One recent example is the noncentrosymmetric superconductor PbTaSe$_2$ with $T_c$ = 3.72 K \cite{Ali}. Angle-resolved photoemission spectroscopy (ARPES) experiments and first-principle calculations revealed topological nodal-line semimetal states and associated surface states in PbTaSe$_2$ \cite{Bian,Chang}.  The thermal conductivity, specific heat, and London penetration depth measurements all demonstrated a fully gapped superconducting state in it \cite{WangMX,ZhangCL,PangGM}, which is also required for a TSC.

More recently, it was argued that the cubic Laves phase Au$_2$Pb ($T_c$ $\approx$ 1.2 K) may be another natural TSC candidate \cite{Cava,Wang4}. Electronic band structure calculations predicted that cubic Au$_2$Pb has a bulk Dirac cone at room temperature \cite{Cava}. With decreasing temperature, Au$_2$Pb undergoes structural phase transitions, and only the orthorhombic phase remains below 40 K \cite{Cava,Baumbach}. Their calculations showed that the structure transition gaps out the Dirac spectrum in the high temperature phase, and results in a low temperature nontrivial massive 3D Dirac phase with Z$_2$ = -1 topology \cite{Cava}. In Ref. \cite{Wang4}, the first-principles calculations also point to the nontrivial topology of the orbital texture near the dominant Fermi surfaces, which suggests the possibility of topological superconductivity. To check whether Au$_2$Pb is indeed a TSC, it will be very important to determine its superconducting gap structure first.

In this letter, we investigate the superconducting gap characteristics of the Au$_2$Pb single crystal, by means of the ultra-low-temperature specific heat and thermal conductivity measurements. Our analysis shows that the electronic specific heat can be described by the two-band $s$-wave model. Furthermore, a negligible $\kappa_0/T$ in zero field and a slow field dependence of $\kappa_0/T$ at low field are revealed by the thermal conductivity measurements. These results suggest that Au$_2$Pb has a fully gapped $s$-wave superconducting state in the bulk.

\section{Experiments}

Single crystals of Au$_2$Pb were grown by a self-flux method \cite{Wang4}. The sample for transport measurements was cut to a rectangular shape of dimensions 2.5 $\times$ 1.0 mm$^2$ in the $ab$ plane, with a thickness of 0.10 mm along the $c$ axis. Contacts were made directly on the sample surfaces with silver paint, which were used for both resistivity and thermal conductivity measurements. The typical contact resistance is less than 100 m$\Omega$ at low temperature. The resistivity was measured in a $^4$He cryostat from 300 K to 2 K, and in a $^3$He cryostat down to 0.3 K. The thermal conductivity was measured in a dilution refrigerator, using a standard four-wire steady-state method with two RuO$_2$ chip thermometers, calibrated $in$ $situ$ against a reference RuO$_2$ thermometer. Magnetic fields were applied along the $c$ axis and perpendicular to the heat current. To ensure a homogeneous field distribution in the sample, all fields were applied at a temperature above $T_c$. The low-temperature specific heat was measured from 0.1 to 3.5 K in a physical property measurement system (PPMS, Quantum Design) equipped with a small dilution refrigerator.

\section{Results and discussion}

\begin{figure}
\includegraphics[clip,width=8cm]{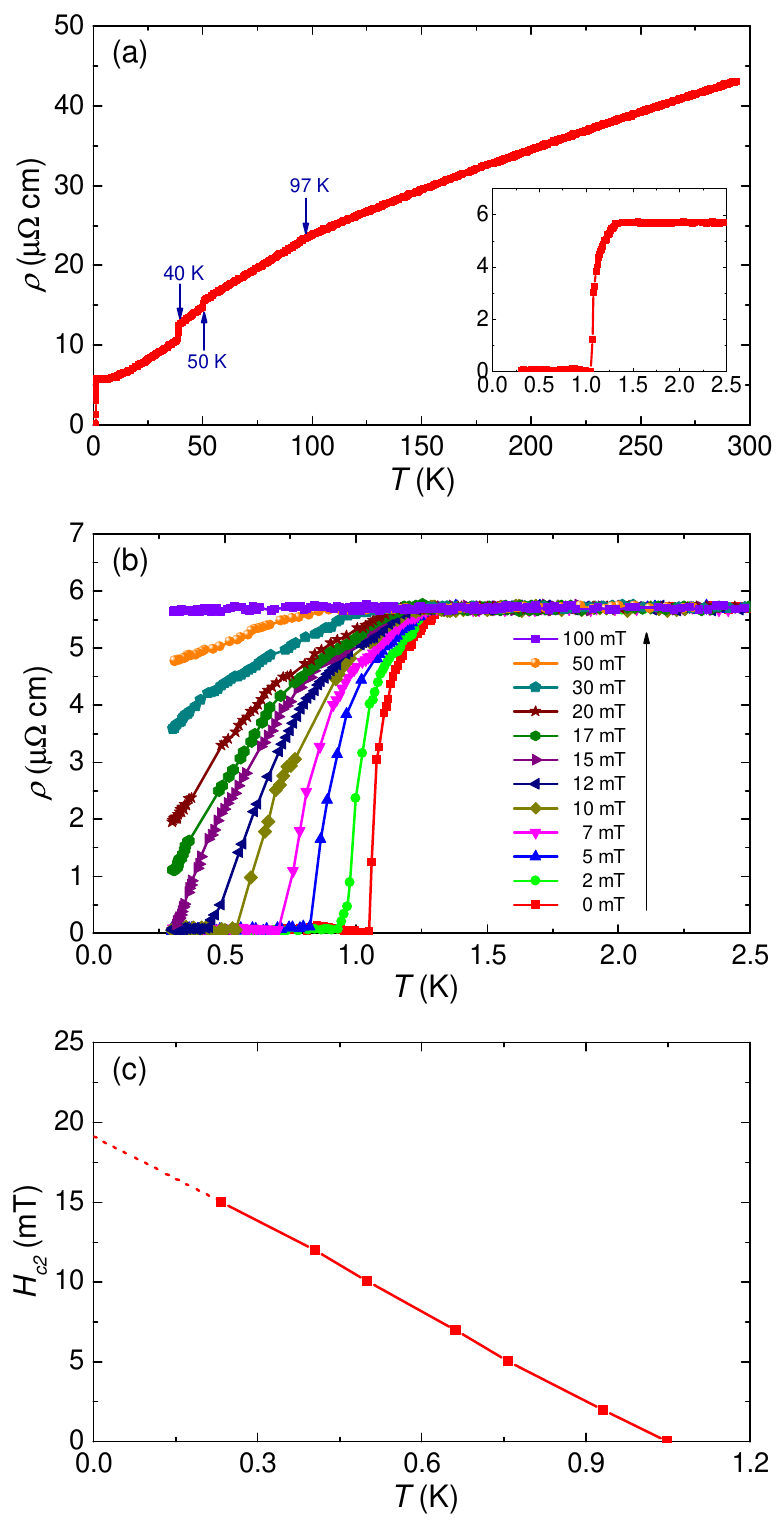}
\caption{(Color online) (a) Temperature dependence of the in-plane resistivity for Au$_2$Pb single crystal in zero field. The arrows indicate the discontinuities related to structural phase transitions. The inset shows the superconducting transition at low temperature. (b) Low-temperature resistivity in magnetic fields up to 100 mT. (c) Temperature dependence of the upper critical field $H_{c2}(T)$, defined by $\rho = 0$. The dashed line is a guide to the eye, from which we estimate $H_{c2}$(0) $\approx$ 19.1 mT.}
\end{figure}

Figure 1(a) shows the in-plane resistivity of the Au$_2$Pb single crystal in zero field. The three discontinuities at 97, 50, and 40 K are related to the structural phase transitions, as previously reported \cite{Cava,Wang4,Baumbach}. From the inset of Fig. 1(a), the width of the resistive superconducting transition (10-90$\%$) is 0.16 K, and the $T_c$ defined by $\rho$ = 0 is 1.05 K. The $\rho(T)$ curve between 1.5 and 2.5 K is quite flat, which extrapolates to a residual resistivity $\rho_0$ = 5.71 $\mu\Omega$ cm. Thus the residual resistivity ratio (RRR = $\rho$(295 K)/$\rho_0$) is about 7.5.

To determine the upper critical field $H_{c2}$(0) of Au$_2$Pb, we measured the low-temperature resistivity of the sample in various magnetic fields up to 100 mT, as shown in Fig. 1(b). The temperature dependence of $H_{c2}(T)$, defined by $\rho$ = 0 on the resistivity curves in Fig. 1(b), is plotted in Fig. 1(c). The dashed line is a guide to the eye, from which we estimate $H_{c2}$(0) $\approx$ 19.1 mT. A slightly different $H_{c2}$(0) does not affect our discussion on the field dependence of $\kappa_0/T$ below. Note that if the $T_{c}$ is defined at the onset of the superconducting transition, the estimated $H_{c2}$(0) will be 86.2 mT, which is comparable to that of Ref. \cite{Cava}.

\begin{figure}
\includegraphics[clip,width=8cm]{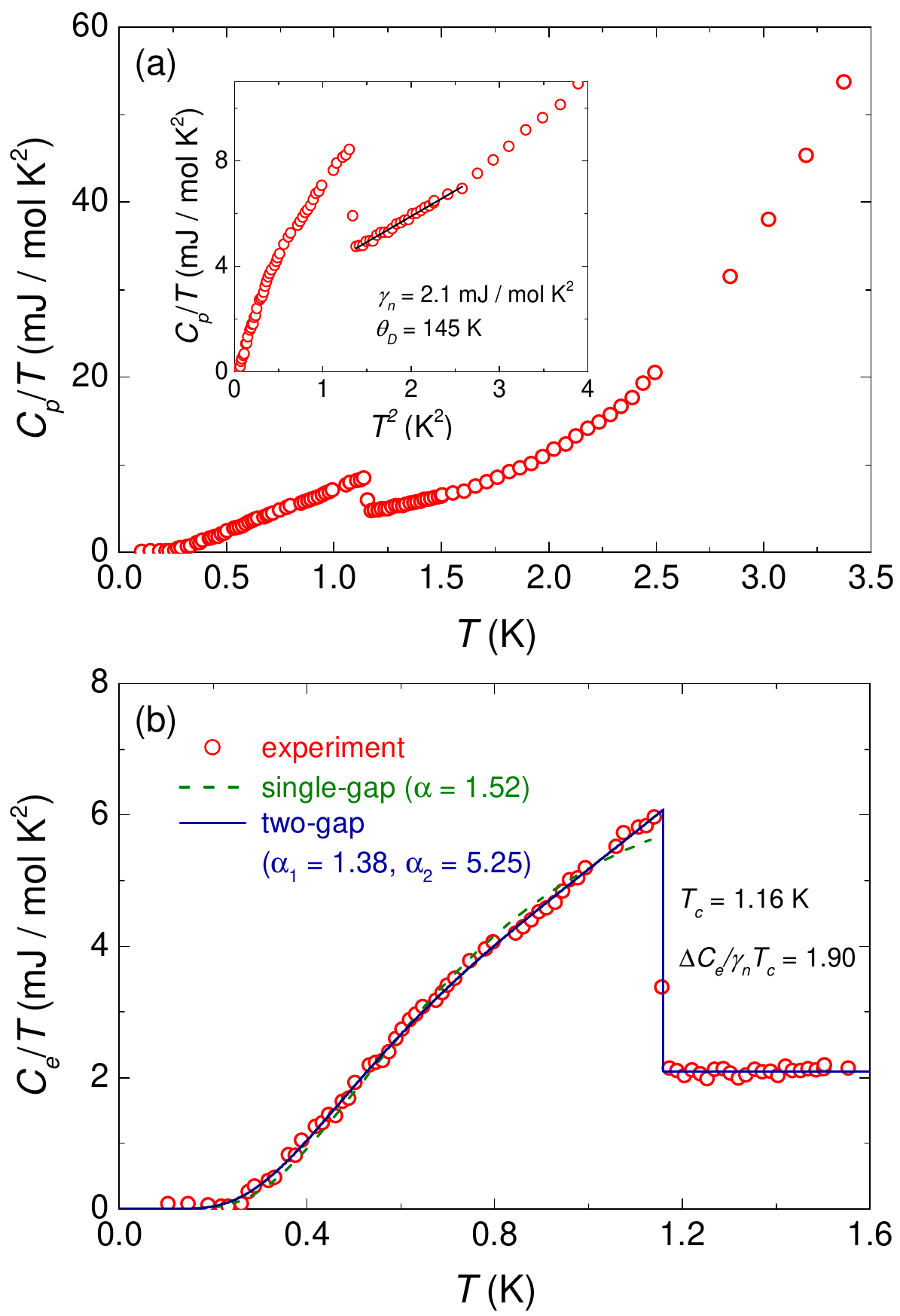}
\caption{(Color online) (a) Temperature dependence of specific heat divided by temperature, $C_p/T$, for Au$_2$Pb single crystal in zero field. The inset shows $C_p/T$ vs $T^2$. The straight line represents the linear fit to $C_p/T = \gamma_n + \beta T^2$ between 1.4 to 2.6 K$^2$. (b) Electronic specific heat $C_e/T$ as a function of temperature $T$. The dashed and solid lines represent the fits using single-band and two-band $s$-wave models, respectively.}
\end{figure}

Figure 2(a) shows the low-temperature specific heat of Au$_2$Pb single crystal down to 0.1 K in zero field. A significant jump due to the superconducting transition is observed at $T_c \approx 1.16$ K, which indicates the high quality of our sample. At low temperature, the normal-state specific heat $C_p$ can be described by $C_p = C_{en} + C_{lattice}$ with the electronic contribution of $C_{en} = \gamma_nT$ and the lattice contribution of $C_{lattice} = \beta T^3$. In the inset of Fig. 2(a), the straight line represents the linear fit of $C_p/T$ from 1.4 to 2.6 K$^2$, yielding $\gamma_n$ = 2.1 mJ mol$^{-1}$ K$^{-2}$, and $\beta$ = 1.9 mJ mol$^{-1}$ K$^{-4}$. From the relation $\theta_D = (12\pi^4RZ/5\beta)^{1/3}$, where $R$ is the molar gas constant and $Z$ is the total number of atoms in one unit cell, the Debye temperature $\theta_D$ = 145 K is estimated. These values are comparable to those previously reported \cite{Cava,Baumbach}.

Figure 2(b) displays the low-temperature electron specific heat, plotted as $C_e/T$ versus $T$. At the superconducting transition, the specific heat jump $\Delta C_e/\gamma_nT_c$ is estimated to be about 1.90, which is higher than the weak-coupling BCS prediction of 1.43. Again, this value is close to that of Ref. \cite{Cava}. In Ref. \cite{Cava}, the specific heat of Au$_2$Pb was measured down to 0.4 K. Here, we measured it down to lower temperature ($\sim$ 0.1 K), which enables us to quantitatively analyze its behavior. We first fit $C_e/T$ in the superconducting state with the BCS $\alpha$-model $C_{es} = C_0$exp$(- \Delta/k_BT)$, where $\Delta$ is the magnitude of the superconducting gap. From Fig. 2(b), the single-band $s$-wave model with $\alpha \equiv \Delta/k_BT_c = 1.52$ cannot describe the experimental data very well. Then we consider the possibility of two gaps, using the phenomenological two-band $\alpha$-model \cite{Phillips,Shiffman}. In this fit, the specific heat is calculated as the sum of the contributions from two bands by assuming independent BCS temperature dependence of the two $s$-wave superconducting gaps. The magnitudes of the two gaps at the $T$ = 0 limit are introduced as adjustable parameters, $\alpha_1 = \Delta_1$(0)/$k_BT_c$ and $\alpha_2 = \Delta_2$(0)/$k_BT_c$, together with the quantity $\gamma_i/\gamma_n$ ($i$ = 1, 2), which is the weight of the total electron density of states (EDOS) for each band. As shown in Fig. 2(b), the two-band fit with $\alpha_1$ = 1.38 and $\alpha_2$ = 5.25 reproduces the specific heat very well. The superconducting gap ratio $\Delta_1$(0)/$\Delta_2$(0) $\approx$ 0.26 is obtained. The weights are 8\% and 92\% for the two bands with the small and large gaps, respectively. This specific heat result indicates a fully gapped superconducting state in Au$_2$Pb.

\begin{figure}
\includegraphics[clip,width=8cm]{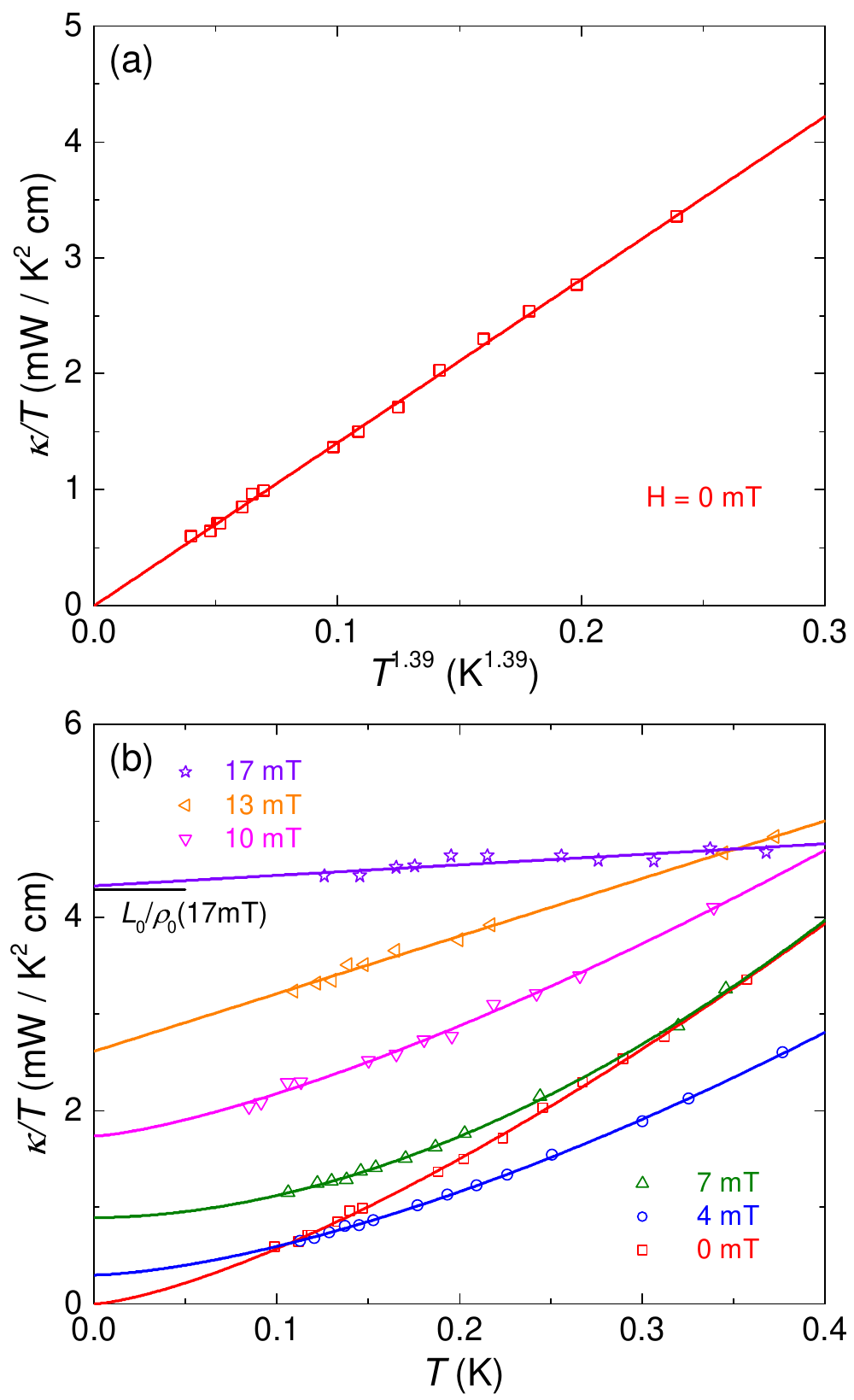}
\caption{(Color online) (a) Temperature dependence of the in-plane thermal conductivity for Au$_2$Pb single crystal in zero field. The solid line represents a fit of the data to $\kappa/T$ = $a$ + $bT^{\alpha-1}$, which gives the residual linear term $\kappa_0/T$ = - 2.6 $\pm$ 6 $\mu$W K$^{-2}$ cm$^{-1}$. (b) Low-temperature thermal conductivity of Au$_2$Pb single crystal in magnetic fields applied perpendicular to the $ab$ plane. The black line is the normal-state Wiedemann-Franz law expectation $L_0/\rho_0$(17 mT), with the Lorenz number $L_0$ = 2.45 $\times$ 10$^{-8}$ W $\Omega$ K$^{-2}$ and $\rho_0$(17 mT) = 5.71 $\mu\Omega$ cm.}
\end{figure}

The ultra-low-temperature thermal conductivity measurement is another bulk technique to probe the superconducting gap structure \cite{Shakeripour}. In Fig. 3, we present the temperature dependence of the in-plane thermal conductivity for Au$_2$Pb single crystal in zero and magnetic fields. The thermal conductivity at very low temperature usually can be fitted to $\kappa/T$ = $a$ + $bT^{\alpha-1}$ \cite{Takagi,Taillefer}, where the two terms $aT$ and $bT^{\alpha}$ represent contributions from electrons and phonons, respectively. In order to obtain the residual linear term $\kappa_0/T$ contributed by electrons, we extrapolate $\kappa/T$ to $T$ = 0 K. Because of the specular reflections of phonons at the sample surfaces, the power $\alpha$ in the second term is typically between 2 and 3 \cite{Takagi,Taillefer}.

We first examine the accuracy of the Wiedemman-Franz law in the normal state of Au$_2$Pb. In Fig. 3(b), the fit of the data in $H_{c2}$ = 17 mT gives $\kappa_{N0}/T$ = 4.33 mW K$^{-2}$ cm$^{-1}$. This value meets the Wiedemann-Franz law expectation $L_0/\rho_0$(17 mT) = 4.29 mW K$^{-2}$ cm$^{-1}$ nicely, with the Lorenz number $L_0$ = 2.45 $\times$ 10$^{-8}$ W $\Omega$ K$^{-2}$ and $\rho_0$(17 mT) = 5.71 $\mu\Omega$ cm. Here we take $H$ = 17 mT as the bulk $H_{c2}$(0) of Au$_2$Pb, which is slightly lower than that estimated from resistivity measurements. The verification of the Wiedemann-Franz law in the normal state shows the reliability of our thermal conductivity measurements.

In zero field, the fitting gives a residual linear term $\kappa_0/T$ = - 2.6 $\pm$ 6 $\mu$W K$^{-2}$ cm$^{-1}$, as seen in Fig. 3(a). Considering our experimental uncertainty $\pm$ 5 $\mu$W K$^{-2}$ cm$^{-1}$, the $\kappa_0/T$ is essentially zero. For nodeless superconductors, there are no fermionic quasiparticles to conduct heat as $T \rightarrow$ 0, since all electrons become Cooper pairs. Therefore, there is no residual linear term $\kappa_0/T$, as seen in conventional $s$-wave superconductors Nb and InBi \cite{Sousa,Ginsberg}. However, for unconventional superconductors with nodes in the superconducting gap, the nodal quasiparticles will contribute a substantial $\kappa_0/T$ in zero field. For example, $\kappa_0/T$ = 1.41 mW K$^{-2}$ cm$^{-1}$ for the overdoped $d$-wave cuprate superconductor Tl$_2$Ba$_2$CuO$_{6+\delta}$ (Tl-2201), which is about 36$\%$ of its $\kappa_{N0}/T$ \cite{Mackenzie}, and $\kappa_0/T$ = 17 mW K$^{-2}$ cm$^{-1}$ for the $p$-wave superconductor Sr$_2$RuO$_4$, which is about 9$\%$ of its $\kappa_{N0}/T$ \cite{Ishiguro}. Therefore, such a negligible $\kappa_0/T$ of Au$_2$Pb in zero field also suggests a fully gapped superconducting state.

\begin{figure}
\includegraphics[clip,width=8cm]{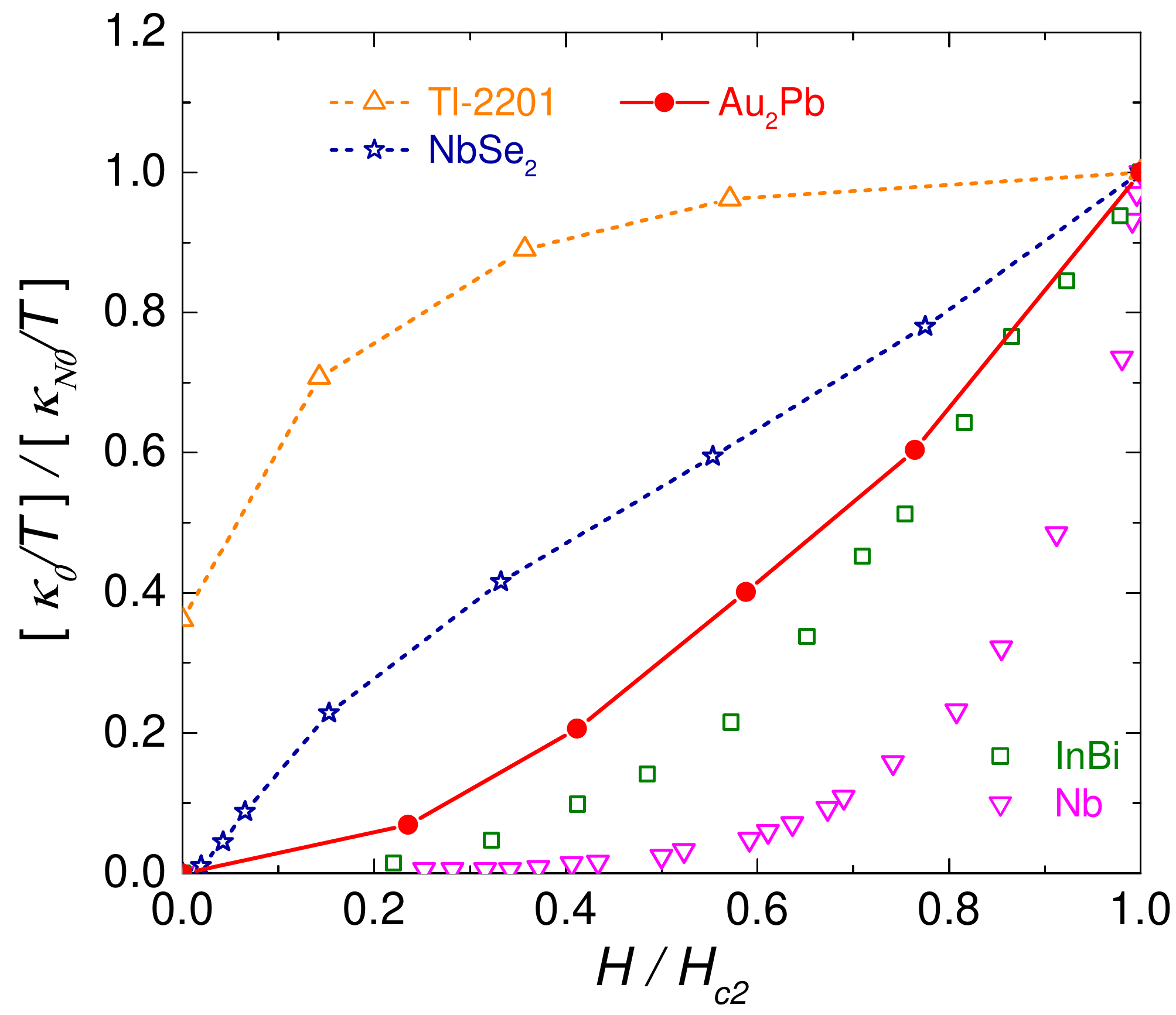}
\caption{(Color online) Normalized residual linear term $\kappa_0/T$ of Au$_2$Pb as a function of $H/H_{c2}$. For comparison, similar data are shown for the clean $s$-wave superconductor Nb \cite{Sousa}, the dirty $s$-wave superconducting alloy InBi \cite{Ginsberg}, the multiband $s$-wave superconductor NbSe$_2$ \cite{Brill}, and an overdoped $d$-wave cuprate superconductor Tl-2201 \cite{Mackenzie}.}
\end{figure}

The field dependence of $\kappa_0/T$ can give more information about the superconducting gap structure \cite{Shakeripour}. Between $H$ = 0 and 17 mT, we fit all the curves and obtain the $\kappa_0/T$ for each magnetic field, as showed in Fig. 3(b). The normalized $\kappa_0/T$ of Au$_2$Pb as a function of $H/H_{c2}$ is plotted in Fig. 4. For comparison, the data of the clean $s$-wave superconductor Nb \cite{Sousa}, the dirty $s$-wave superconducting alloy InBi \cite{Ginsberg}, the multiband $s$-wave superconductor NbSe$_2$ \cite{Brill}, and an overdoped $d$-wave cuprate superconductor Tl-2201 \cite{Mackenzie}, are also plotted. For a clean $s$-wave superconductor with a single gap, $\kappa_0(H)/T$ should grow exponentially with field, as observed in Nb \cite{Sousa}. While for $s$-wave InBi in the dirty limit, the curve is exponential at low $H$, crossing over to a roughly linear behavior closer to $H_{c2}$ \cite{Ginsberg}. In the case of NbSe$_2$, the distinct $\kappa_0(H)/T$ behavior was well explained by multiple superconducting gaps with different magnitudes \cite{Brill}.

In Fig. 4, the field dependence of $\kappa_0(H)/T$ for Au$_2$Pb grows slightly faster than that of the dirty $s$-wave superconductor InBi. Since the residual resistivity $\rho_0$ = 5.71 $\mu\Omega$ cm of our Au$_2$Pb single crystal is quite small, it is unlikely to be a dirty superconductor. We cannot confirm this, since we do not know enough parameters, such as Fermi velocity, to estimate the mean free path of the carriers in Au$_2$Pb. In fact, such a $\kappa_0(H)/T$ behavior may result from multiple superconducting gaps. Electronic structure calculations of Au$_2$Pb showed that there are several bands which cross the Fermi level along the $\Gamma$-$K$ line \cite{Cava}. Analysis on electronic specific heat also reveals two superconducting gaps with the amplitudes $\Delta_1$(0)/$k_BT_c$ = 1.38 and $\Delta_2$(0)/$k_BT_c$ = 5.25, respectively. Note that the weight of the small gap is low (8\%), which may be the reason why the $\kappa_0(H)/T$ of Au$_2$Pb grows much slower than that of NbSe$_2$.

So far, PbTaSe$_2$ and Au$_2$Pb are the only two natural TSC candidates. Our results demonstrate that Au$_2$Pb has similar fully gapped superconducting state to PbTaSe$_2$. While the topological band structure and surface states have been verified in PbTaSe$_2$, detailed ARPES experiments on Au$_2$Pb are highly desired. Note that recent scanning tunneling microscopy (STM) experiments on PbTaSe$_2$ observed a full superconducting gap in zero field, and the zero energy bound states at superconducting vortex cores in magnetic field \cite{Guan}. However, the thermal broadening at $T$ = 0.26 K makes it impossible to discriminate the ordinary Caroli-de Gennes¨CMatricon (CdGM) bound states and Majorana bound states \cite{Guan}. Therefore, more STM experiments, such as spin-polarized STM \cite{Sun}, are needed to determine whether these two natural TSC candidates are indeed TSC.

\section{Summary}

In summary, the superconducting gap structure of Au$_2$Pb has been studied by the ultra-low-temperature specific heat and thermal conductivity experiments. The analysis of the electronic specific heat suggests two $s$-wave superconducting gaps with the amplitudes $\Delta_1$(0)/$k_BT_c$ = 1.38 and $\Delta_2$(0)/$k_BT_c$ = 5.25. Furthermore, a negligible $\kappa_0/T$ in zero field and the field dependence of $\kappa_0/T$ also suggest nodeless superconducting gaps. These results indicate that the second natural TSC candidate Au$_2$Pb has a fully gapped superconducting state in the bulk.

\acknowledgments
This work is supported by the Natural Science Foundation of China, the Ministry of Science and Technology of China (National Basic Research Program No. 2012CB821402 and No. 2015CB921401), Program for Professor of Special Appointment (Eastern Scholar) at Shanghai Institutions of Higher Learning, and STCSM of China (No. 15XD1500200). The work in Peking University was supported by National Basic Research Program of China (Grant Nos. 2013CB934600 and 2012CB921300), the Open Project Program of the Pulsed High Magnetic Field Facility (Grant No. PHMFF2015002), Huazhong University of Science and Technology, and Open Research Fund Program of the State Key Laboratory of Low-Dimensional Quantum Physics.\\

\end{document}